\documentstyle[12pt,aasms4,epsf]{article}

\def\etal   {{\rm ~et al.,~}}
\def\kms    {\ifmmode{{\rm ~km~s}^{-1}}\else{~km~s$^{-1}$}\fi}
\def\lsun   {\ifmmode{{\rm ~L}_\odot}\else{~L$_\odot$}\fi}
\def\msun   {\ifmmode{{\rm ~M}_\odot}\else{~L$_\odot$}\fi}

\slugcomment{\it To appear in Ap.J. (Letters)}

\lefthead{Greenhill et al.}
\righthead{Water Maser Emission in NGC\,4945}

\begin{document}

\title{The Distribution of H$_2$O Maser Emission in the Nucleus of NGC\,4945}
\author{L. J. Greenhill, J. M. Moran, \& J. R. Herrnstein\altaffilmark{1}}
\affil{Harvard-Smithsonian Center for Astrophysics, 60 Garden St, 
Cambridge, MA 02138}
\altaffiltext{1}{Currently at the National Radio Astronomy Observatory,
Socorro, NM 87801}


\begin{abstract}

We present the first interferometer map of the H$_2$O maser emission in the active
nucleus of NGC\,4945, which exhibits both starburst and Seyfert qualities.
Although the declination of the galaxy is $-49^\circ$, we were able to make 
the observations with the southernmost antennas of the Very Long Baseline Array.
Strong maser emission is present in three velocity ranges, one near the systemic
velocity and two shifted roughly symmetrically by $\pm (100-150)\kms$. This is the
first detection of highly blue-shifted water emission in NGC\,4945. We also report
a marginal detection of previously unreported red-shifted emission at a 
velocity of $\sim +200 \kms$
with respect to systemic.
From fringe rate analysis we determined the position of the maser to be
$\alpha_{1950}=13^h02^m32\rlap{.}^s28\pm0\rlap{.}^s02 ; \delta_{1950}=-49^\circ
12'01\rlap{.}''9\pm0\rlap{.}''1.$ The uncertainties in earlier estimates are at
least several arcseconds. The maser lies within $2''$ (36 pc at a distance of 3.7
Mpc) of the peaks in 1.4 GHz continuum and 1.6 $\mu$m emission from the nucleus.
The mappable maser emission is distributed roughly linearly over
$\approx 40$ milliarcseconds (0.7 pc) at a position angle of $\approx 45^\circ$,
which is close to the $43\pm 2^\circ$ position angle of the galactic disk.
The red and blue-shifted emission symmetrically stradle the systemic emission on the
sky, which suggests material in edge-on circular motion around a central object.
The position-velocity structure indicates a binding
mass of $\sim 1\times10^6$ M$_\odot$, within a volume of radius $\approx 0.3$ pc, which in
combination with estimates of the AGN luminosity, implies that the central engine
radiates on the order of 10\% of its Eddington luminosity.

\end{abstract}


\section{Introduction}

The edge-on spiral galaxy NGC\,4945 contains the first H$_2$O maser discovered
in a galactic nucleus (\cite{DL79}; \cite{B82}). Within the nucleus there is a
heavily absorbed hard X-ray component ($>10$ keV), with a measured column
density of $n_H=5\times10^{24}$ cm$^{-2}$ (\cite{Iwa93}), and true X-ray
luminosity (2--20 keV) of $9\times10^{41}$ erg s$^{-1}$. The hard component
varies on timescales of hours, implying a source size of $\sim 10^{15}$ cm.
The nucleus is very prominent at infrared wavelengths, with a far infrared
luminosity $L_{\rm FIR}=2\times10^{43}$ erg s$^{-1}$ from the inner 200 pc 
(\cite{Brock88}), at
an assumed distance of 3.7 Mpc (\cite{WW90}). 
The nucleus also shows a $\sim 1\times6$ kpc
ionization cone along the galactic minor axis (\cite{Nak89}) and emission knots
with LINER-like spectra (\cite{Moor95}). Optical line splittings of
up to $600\kms$~(\cite{HAM90}, \cite{Moor95}) indicate that the cone is
evacuated by a nuclear wind. While X-ray characteristics suggest that the active
galactic nucleus (AGN) is driven by an obscured Seyfert nucleus, the high IR
luminosity and wind indicate the presence of a nuclear starburst.

Spectra of the maser consistently show emission ($\sim 1$ Jy) within about
$\pm20\kms$~of the systemic velocity and stronger emission  ($\sim 5-10$ Jy)
red-shifted by $\sim 100-150\kms$ (e.g., \cite{WG86}, Figure~1). Previously, no 
corresponding emission
on the blue side of the systemic velocity has been detected 
(e.g., \cite{Nak95}). We adopt a heliocentric systemic velocity of $561\pm 4$\kms, based on 
CO observations (\cite{D93}).
(Velocities are defined with respect to the
standard radio astronomical definition of Doppler shift throughout.) The importance of
nuclear H$_2$O masers to the study of AGN has been widely appreciated after
recent observations of the maser in the nucleus of NGC 4258 (\cite{Miy95}),
where the masers were found to trace a thin molecular Keplerian disk of 0.13
pc inner radius around a central mass of $3.6\times10^7$ M$_\odot$. In
this Letter, we report the first sub-arcsecond estimate of the maser position
in NGC\,4945, the detection of blue-shifted maser features,  
and the first interferometric map of the distribution of maser emission.

\section{Observations \& Results} 

We observed the maser with the Very Long Baseline Array of the NRAO$^1$ for 4
hours 

\noindent
\hrulefill

\noindent
$^1$ The National
Radio Astronomy Observatory is operated by Associated Universities, Inc, under
cooperative agreement with the National Science Foundation.

\noindent
\hrulefill

\noindent
on 1995 January 22. We used the southernmost stations, at Mauna Kea, HI
(MK), Saint Croix, VI (SC),
and Kitt Peak, AZ (KP). At KP the source reached a maximum elevation of 
$8.6^\circ$, while it was $11^\circ$ and rising at MK.  We obtained 
useful data for only 2.5 hours on the MK--KP baseline. 
Observing conditions at Saint Croix were relatively poor and no useful
data were obtained from it.  Tipping scans
showed that the opacity at KP, MK, and SC were 0.07, 0.05, and 0.3 
at zenith, respectively.  The baseline length was 4470 km and 
the angular resolutions were $\sim0.6$ and $\sim 1.8$ milliarcseconds (mas) in right ascension 
and declination, respectively.  The system temperatures for $8^\circ$ elevation at
KP and MK were 170 and 130 K, respectively. 
We recorded four 
contiguous bandpasses covering the range 405\kms~to 821\kms, 
so as to include the systemic
emission, the red-shifted emission, and the likely velocity range of counterpart blue-shifted 
emission. The spectrum is shown in Figure~1.

We calibrated the data with the NRAO AIPS package. The interferometer delays and
fringe rates were estimated from observations of the continuum sources 4C39.25,
3C273, and $1147-382$. The latter source was observed at elevations $< 20^\circ$
and was included 
so that we could estimate the magnitude of elevation dependent 
errors in the correlator model and post-correlation analysis.
The delay calibration is accurate to about 3 ns overall, after
implementing the U. S. Naval Observatory station positions and earth orientation
parameters (Eubanks, private communication).

We analyzed the fringe rate of the maser emission as a function of time to
estimate the position of the maser, taking into account the effect of the
atmosphere. We fit a sinusoid to the measured fringe rates for emission between
664 and 719\kms~(with uniform weighting) to obtain position offsets of $+2.7''$
and $+8.2''$ from the {\it a priori} position
$\alpha_{1950}=13^h02^m32\rlap{.}^s000$, $\delta_{1950}=-49^\circ
12'10\rlap{.}''05$, with respect to a reference frame consisting of the sources
4C39.25, $1147-382$, and 3C273B. We conservatively estimate the uncertainties in
position based on the residual fringe rates of the maser and calibrator sources.
The new maser position is

{
\parskip=0pt

$$\alpha_{1950}=13^h02^m32\rlap{.}^s28\pm0\rlap{.}^s02 , 
\delta_{1950}=-49^\circ 12'01'\rlap{.}''9\pm0\rlap{.}''1; $$
$$\alpha_{2000}=13^h05^m27\rlap{.}^s48\pm0\rlap{.}^s02  , \delta_{2000}=-49^\circ 
28'05\rlap{.}''4\pm0\rlap{.}''1.$$

}

To map the brightness distribution of the maser, we phase referenced the data to
the strong spectral feature at $699\pm 0.8\kms$. The phases of this feature were
subtracted from the phases of other spectral features, which corrected for the
effects of the atmosphere and calibration uncertainties. We assumed a point
source model and fit a sinusoid (of 24 hour period) to the difference phase,
estimating relative feature positions and formal errors (e.g., \cite{TMS86}).
Sample phase plots are shown in Figure~2. The data were averaged in frequency
across the line profile for each feature and averaged in time for between 300 s
and 640 s so that phase ambiguities could be resolved prior to fitting. For
a 1 Jy feature that is 1\kms wide (half power 
width) the signal-to-noise ratio in 300 s is about 4.

The structure of the maser is remarkably linear, with emission distributed over
$\approx 40$ mas between 414 and 719\kms (Figure~3). Deviations from linearity
are
antisymmetric with respect to the systemic emission, which gives the apprearance
of a shallow ``S'' on the sky.  The centroids of the blue and red emission are
separated by $\approx 30$ mas or 0.55 pc, at a position angle of $\approx
45^\circ$, and roughly symmetrically bracket the systemic emission. The wide
separations among the systemic, red, and blue emission correspond to many
interferometer fringes and are well determined. Maser features that are detected
only marginally (i.e., with a rms of residuals about the fitted model of $> 1$
rad) are not shown in the map, although they do not appreciably alter the source
structure. These marginal features include one at $\approx 779\kms$, which is
more red-shifted than any previously reported maser emission in NGC\,4945. In
Nonetheless, not all of the detected maser features could be mapped, particularly
near the systemic and blue-shifted velocities, because a significant fraction
are not well approximated by a point source model (i.e., the phases could not be
fit by a diurnal sinusoid).

The interferometer calibration was sufficiently good that uncertainty in the
astrometric position of the maser is the largest source of quantifiable
systematic error in estimating maser feature positions. The error is $\la 90$
$\mu$as, which is computed for an astrometric uncertainty of 100 mas and for a
feature 200\kms~from the reference, and is proportional to both quantities.
(Residual atmospheric delay contributes a moderate systematic error because 
of the low
elevations at which we observed the maser; it is smaller than the error from
the astrometry.) A second source of systematic error arises from the possibility
that emission, for each feature,  may not be well described by a point source,
which is difficult to quantify.  However, the data for each mapped maser feature
is well fit by a point source model over the limited range of hour angle
available. The systematic error in position for each maser feature is less than
the statistical error, which is dominated by the limited range of hour angle
over which the phases were measured. This limitation leads to 1) a large
correlation between right ascension and declination parameters and 2) an
uncertainty as to in which lobe of the interferometer fringe pattern the maser
lies. Uncertainties in the estimated positions take into account possible lobe
ambiguities.

\section{Discussion}

The new astrometric position of the maser clearly shows that it is
associated with the center of the galactic nucleus (Figure~5).  The previous
position of Batchelor\etal (1982), obtained with the Parkes antenna of the
CSIRO, was substantially more uncertain because of the beamwidth of the
antenna. The maser lies within $1\sigma$ of the infrared peak at 1.6 $\mu$m
and overlaps the radio continuum source, which Whiteoak \& Wilson (1990) estimate
is $6''\times2''$ in size at 6 GHz (see Figure~5).

We cautiously interpret the data for the maser in NGC\,4945 in terms of the
paradigm established for NGC~4258. The extensive data on NGC~4258, with its
clear Keplerian signature, show that the masers define a nearly edge on thin
annular disk. The data for NGC4945 are more sketchy but they share the
characteristic of systemic emission bracketted roughly symmetrically in velocity
{\it and} position by high-velocity emission. Moreover, the angular
distributions of blue-shifted emission is roughly linear and the velocities
decline with angular distance from the systemic emission. However, the position
errors are significant, the red-shifted emission does not appear to mirror this
structure, the decline in velocity for the blue-shifted emission is faster than
what is characteristic for a Keplerian rotation curve, and the non-point source
emission that seems to exist at some systemic and blue-shifted velocities could not be
mapped (Figure~4).  Nonetheless, there is a suggestion that the structures of NGC\,4945 and
NGC\,4258 may be similar, and we proceed on that assumption.

In the context of an edge-on disk, the observations suggest a binding mass (for
circular orbits) of $M = 1.1 V^2 D_{\rm Mpc} b_{\rm mas} \csc^2 i \csc^3\theta$,
where $M$ is the enclosed mass in units of M$_\odot$, $V$ is the orbital
velocity in units of \kms, $D_{\rm Mpc}$ is the distance in Mpc, $b_{\rm mas}$
is the radius in mas, $i$ is inclination, and $\theta$ is the azimuthal angle of
the red-shifted masers in the disk plane ($90^\circ$ is the plane of the sky).
The binding mass is $M \approx 1.4\times10^6$ M$_\odot$ for $V=150\kms$, $b=15$
mas or 0.3 pc, and $i=\theta=90^\circ$ (Figure~4). The implied mass density is
$\sim 1.6\times10^7$ M$_\odot$ pc$^{-3}$. 
The position angle of the disk, $\approx 45^\circ$, is comparable to 
the position angle of
the larger galactic disk, $43\pm2^\circ$, and the senses of rotation of the two
disks are the same.  
If the observed $9\times10^{41}$ erg s$^{-1}$ X-ray
luminosity (\cite{Iwa93}) implies a bolometric luminosity on the order of
$10^{43}$ erg s$^{-1}$, then the central engine radiates 10\% of its Eddington
luminosity. We note that the shallow ``S'' the mappable maser emission 
traces may be consistent with a warp in the disk, which could be caused by
radiation-driven torques that result from oblique illumination 
by the central engine (\cite{Mal96}), though this is not a definitive explanation.

The interpretation of the observed maser structure is not unique and radically
different interpretations might be viable. The emission could arise from a
complex of star forming regions superposed on or within the nucleus, or from
jet-induced shocks in the AGN. The linear extent of the maser is comparable to
galactic star formation complexes (e.g., W49N), although the velocity difference
between the systemic and red-shifted features is large for adjacent star forming
regions. Maser emission can arise in warm gas behind dissociative shocks
(\cite{EHM89}) and nondissociative shocks (\cite{KN96}), although the maser
spectrum mandates a complex line-of-sight velocity structure for the shocks.

Interferometric synthesis observations will undoubtedly clarify the dynamical
characteristics of the maser region, providing a comprehensive map of the maser
emission within 200\kms~of the systemic velocity.  The dual nature of
the nucleus, as both a starburst and a very heavily obscured Seyfert object,
makes this water maser source particularly important since it may be used to
probe the dynamics and structure of the nucleus on sub-parsec scales.

\acknowledgments
We thank P. Diamond for his assistance with aspects of the post-correlation 
processing, and S. Ellingsen for assistance in obtaining a spectrum of the maser
at Parkes.

\clearpage

\noindent
Fig. 1-- Spectrum of the water maser emission from the nucleus of NGC\,4945 made
on 1995 October 13 at the Parkes radiotelescope of the CSIRO. The maser is 
variable on time scales of months but the spectrum shown is typical.
The integration
time is about 120 minutes (on and off source). The adopted heliocentric systemic
velocity of the galaxy, 561\kms, is indicated by the vertical bar. The dashed
lines indicate the extremes of velocity that we observed with the VLBA. The
solid bars above the velocity axis indicate the range of velocity within which
maser emission is detected. The absolute gain calibration is accurate to $\sim
30$ \%. Velocities with respect to the Local Standard of Rest may be obtained
from the relation $V_{\rm LSR} = V_{\rm hel} - 4.6\kms$.

\noindent
Fig. 2-- Sample plots of phase vs time for ({\sl top}) the systemic maser
feature at 557\kms~and ({\sl bottom}) the red-shifted feature at 665\kms. The
phases sample a 24 hour sinusoid and are relative to the phase of the reference
feature at 699\kms. The right ascension and declination offsets of the
maser features, with respect to the reference, were determined from these
curves. 
The hour angle of the source with respect to the baseline is $-8.5$ h at
12 UT.

\noindent
Fig. 3-- Map of the maser brightness distribution. The gray-scale shows velocity
(414--719\kms) with lighter shading indicative of more blue-shifted emission. The
error bars reflect measurement noise at the $3\sigma$ level. The position angle
of the galactic disk is $43\pm2^\circ$ (\cite{D93}). 
Marginal detections, for which the rms of the phase residuals from the model are $>1$ rad, 
are also not shown. The arrow indicated the maser feature closest to the 
systemic velocity of the galaxy.

\noindent
Fig. 4-- Position-velocity diagram for the maser emission. The impact parameter
is measured with respect to the maser feature closest to the systemic velocity
of the galaxy. The dashed line indicated the systemic velocity. The error bars 
represent position uncertainties at the $3\sigma$ level. 

\noindent
Fig. 5-- Comparison of various position determinations for the nucleus of
NGC\,4945. These include the new maser position reported here and the
{\it a priori} position obtained by Batchelor\etal (1982) with the Parkes
antenna, the interferometric position of H~I absorption (\cite{Ables87}), the
synthesis positions  at 408 MHz (\cite{Large81}), 1.4 GHz (\cite{Ables87}),
and 6.0 GHz (\cite{WW90}), and the position of the $1.6~\mu$m peak (Whiteoak
\& Gardner, unpublished). The sizes of the symbols correspond to the $1\sigma$
uncertainties in position. The deconvolved angular size of the continuum source at 6 GHz
is about $6''\times2''$ (\cite{WW90}) at a position angle of $\sim 45^\circ$ and is
shown by the dashed ellipse. The position angle of the galactic disk is $43\pm2^\circ$ 
(\cite{D93}). 

\clearpage

\begin{figure}
\plotfiddle{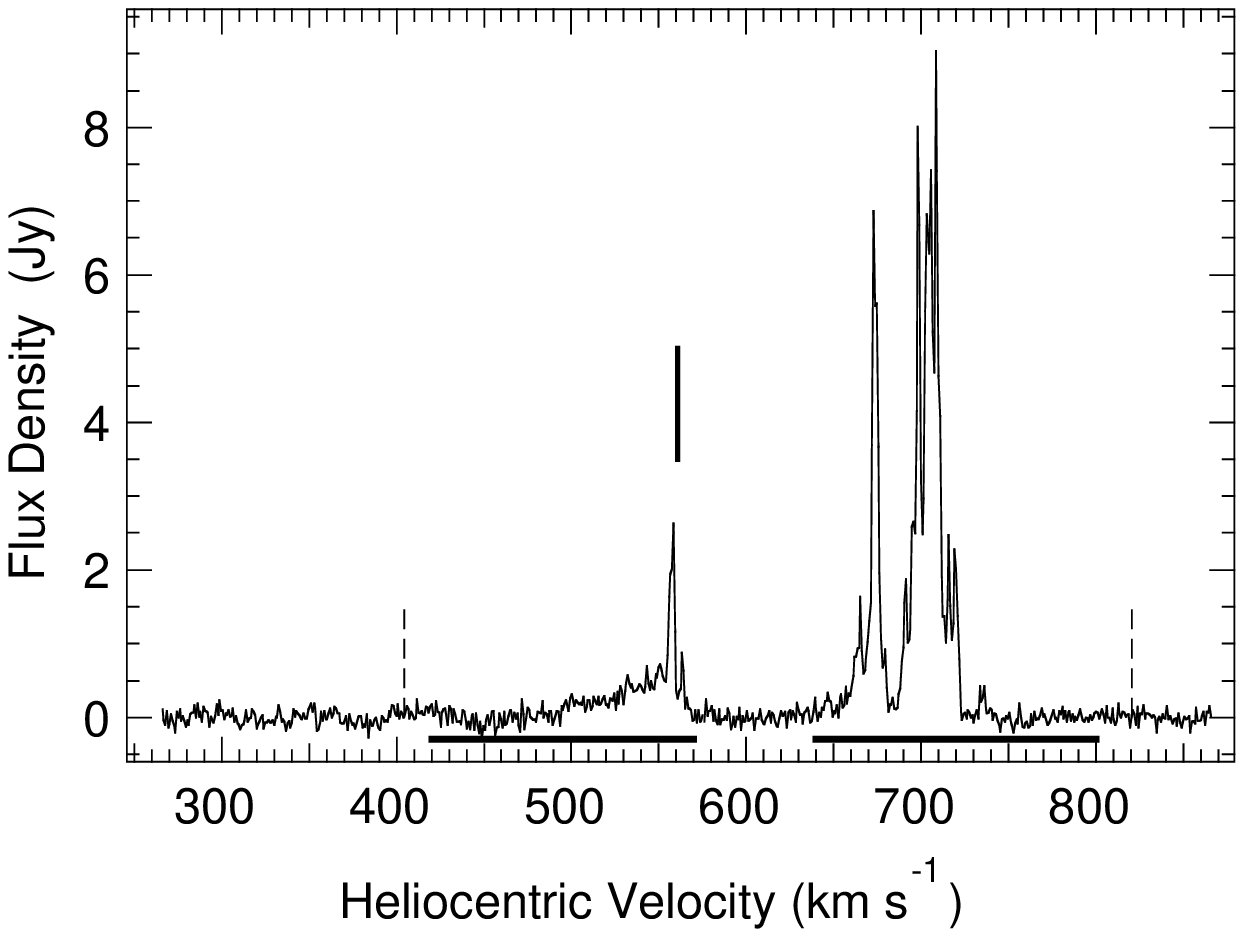}{6.0truein}{0}{100}{100}{-275}{-150}
\caption{Greenhill et al.}
\end{figure}

\clearpage

\begin{figure}
\plotfiddle{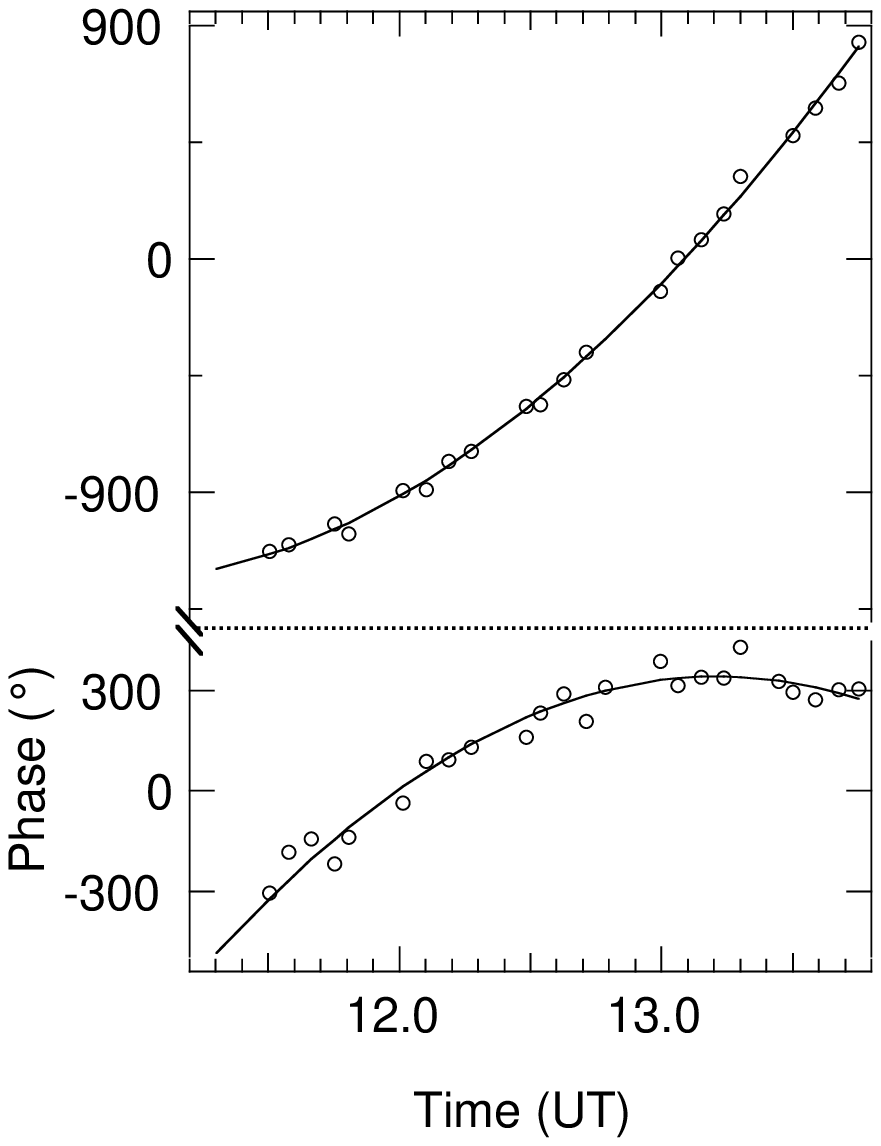}{6.0truein}{0}{100}{100}{-275}{-150}
\caption{Greenhill et al.}
\end{figure}

\clearpage

\begin{figure}
\plotfiddle{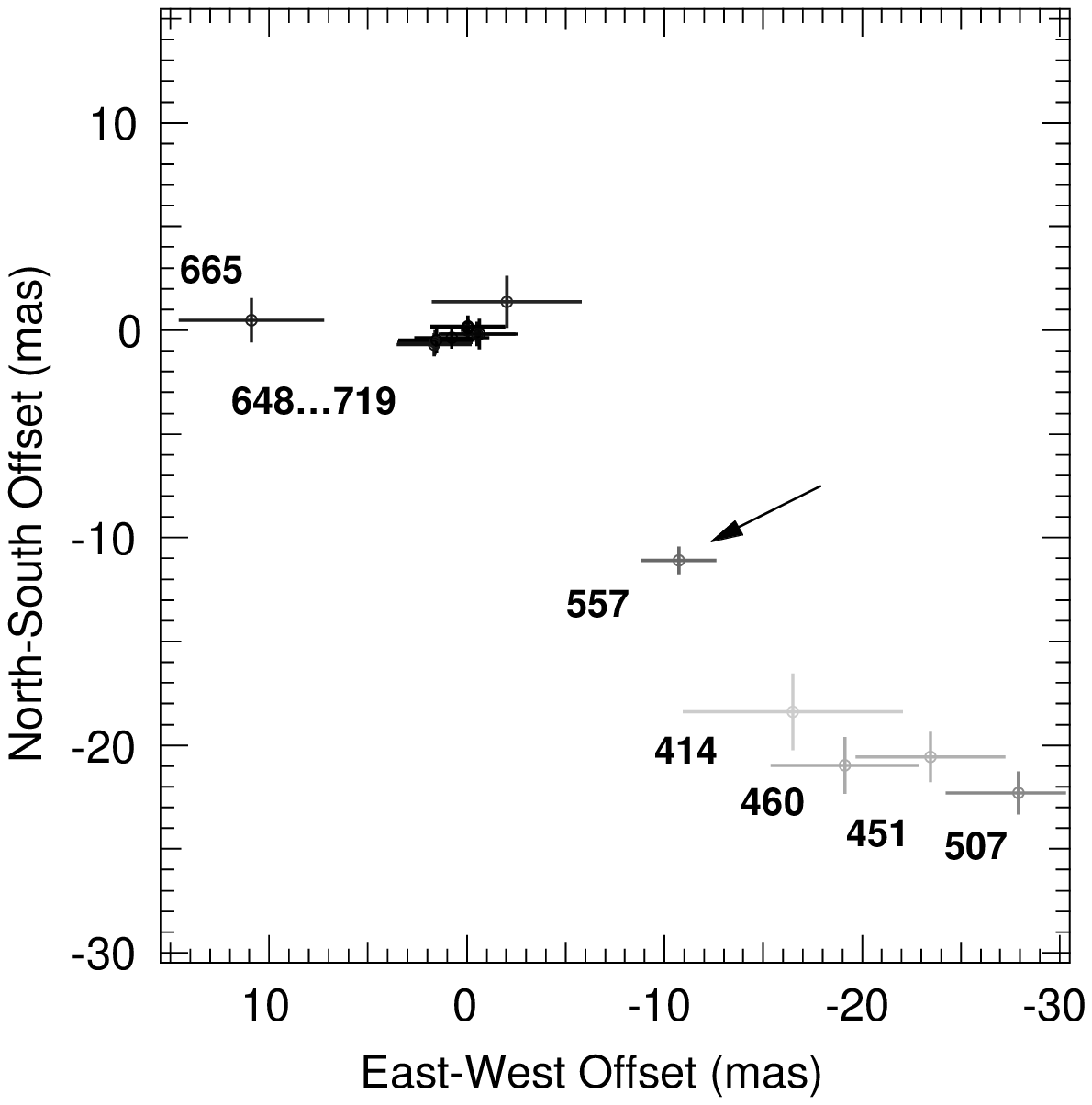}{6.0truein}{0}{100}{100}{-275}{-150}
\caption{Greenhill et al.}
\end{figure}

\clearpage

\begin{figure}
\plotfiddle{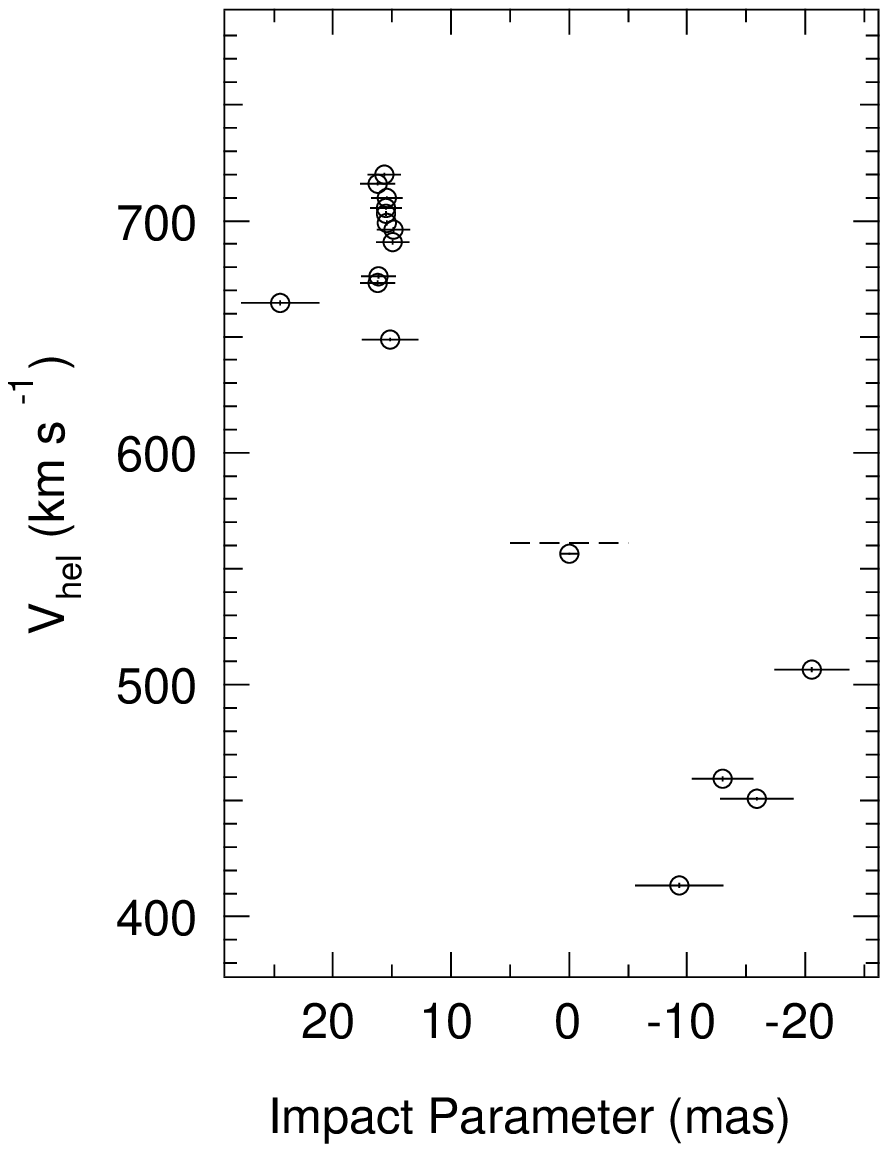}{6.0truein}{0}{100}{100}{-275}{-150}
\caption{Greenhill et al.}
\end{figure}

\clearpage

\begin{figure}
\plotfiddle{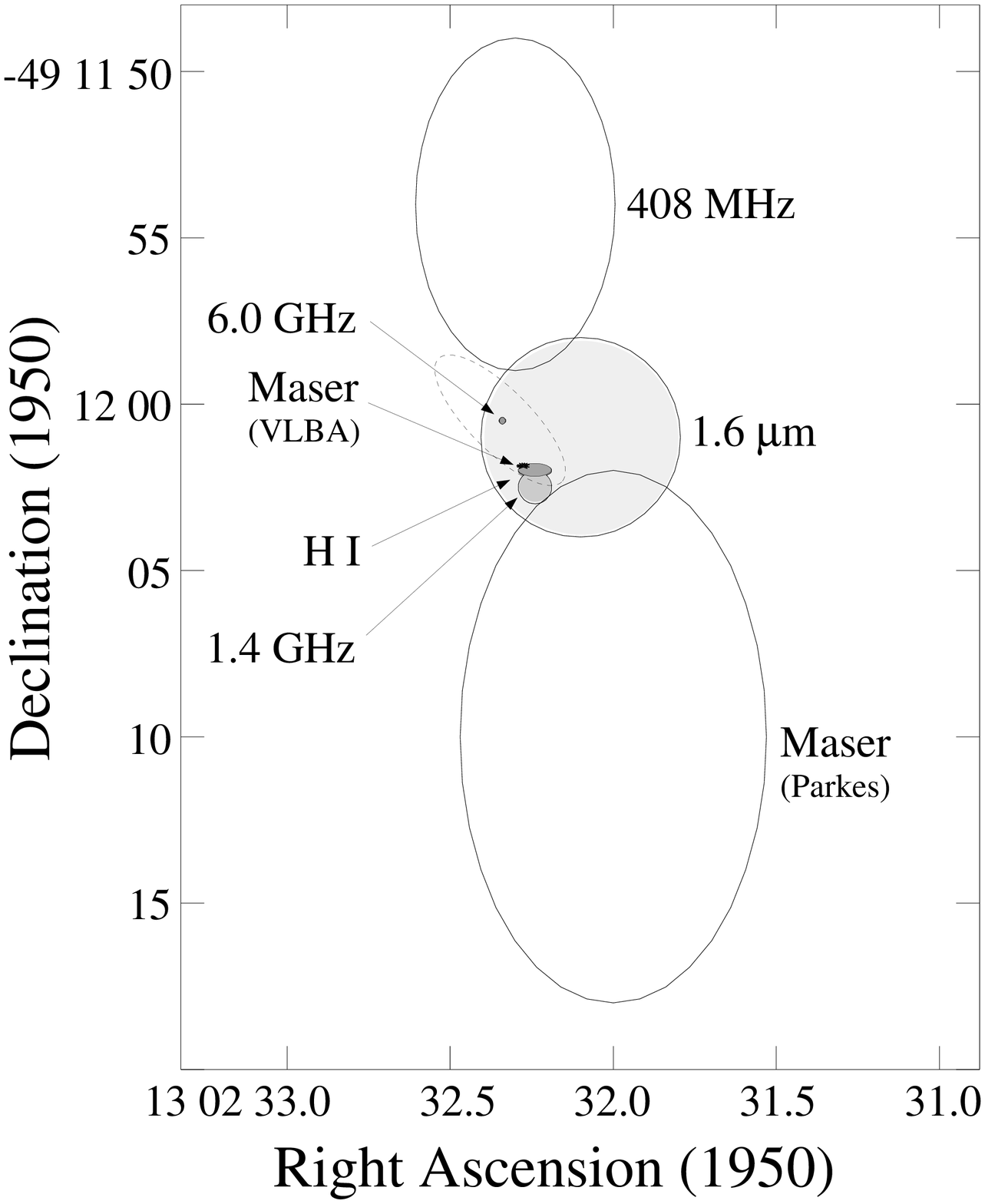}{6.85truein}{0}{90}{90}{-255}{-100}
\caption{Greenhill et al.}
\end{figure}


\begin{thebibliography}{}

\bibitem[Ables\etal 1987]
{Ables87}
Ables, J. G.,\etal 1987, \mnras, 226, 157

\bibitem[Batchelor, Jauncey, \& Whiteoak 1982]
{B82}
Batchelor, R. A., Jauncey, D. A., \& Whiteoak, J. B. 1982, \mnras, 200, 19P

\bibitem[Brock\etal 1988]
{Brock88}
Brock, D., Joy, M., Lester, D. F., Harvey, P. M., \& Ellis, H. B. 1988, \apj
329, 208

\bibitem[Dahlem\etal 1993]
{D93}
Dahlem, M., Golla, G., Whiteoak, J. B., Wielebinski, R., H\"uttemeister, S.,
\& Henkel, C. 1993, A\&A, 270, 29

\bibitem[dos Santos \& L\'epine 1979]
{DL79}
dos Santos, P. M., \& L\'epine, J. R. D. 1979, Nature, 278, 34

\bibitem[Elitzur, Hollenbach, \& McKee 1989]
{EHM89}
Elitzur, M., Hollenbach, D. J., \& McKee, C. F., 1989, \apj, 346, 983

\bibitem[Heckman, Armus, \& Miley 1990]
{HAM90}
Heckman, T. M., Armus, L., \& Miley, G. K. 1990, \apjs, 74, 833


\bibitem[Iwasawa\etal 1993]
{Iwa93}
Iwasawa, K., \etal 1993, \apj, 409, 155

\bibitem[Kaufman \& Neufeld 1996]
{KN96}
Kaufman, M. J., \& Neufeld, D. A. 1996, \apj, 456, 250

\bibitem[Large\etal 1981]
{Large81}
Large, M. I., Mills, B. Y., Little, A. G., Crawford, D. F., \& Sutton, J. M. 1981, \mnras, 194, 693

\bibitem[Maloney, Begelman, \& Pringle 1996]
{Mal96}
Maloney, P. R., Begelman, M. C., \& Pringle, J. E. 1996, ApJ, 472, 582

\bibitem[Miyoshi\etal 1995]
{Miy95}
Miyoshi, M., Moran, J. M., Herrnstein, J. R., Greenhill, L. J., Nakai, N., Diamond, P. J., \& Inoue, M.
1995, Nature 373, 127

\bibitem[Moorwood\etal 1995]
{Moor95}
Moorwood, A. F. M. van der Werf, P. P., Kotilainen, J. K., Marconi, A., \& Oliva, E. 
1995, \aap, 308, L1

\bibitem[Nakai 1989]
{Nak89}
Nakai, N. 1989, PASJ, 41, 1107

\bibitem[Nakai\etal 1995]
{Nak95}
Nakai, N., Inoue, M., Miyazawa, K., Miyoshi, M., \& Hall, P. 1995, PASJ, 47, 771

\bibitem[Thompson, Moran, \& Swenson 1986]
{TMS86}
Thompson, A. R., Moran, J. M., \& Swenson, G. W.  1986, 
Interferometry and Synthesis in Radio Astronomy, (New York: Wiley)



\bibitem[Whiteoak \& Gardner 1986]
{WG86}
Whiteoak, J. B., \& Gardner, F. F. 1986, MNRAS, 222, 513

\bibitem[Whiteoak \& Wilson 1990]
{WW90}
Whiteoak, J. B., \& Wilson, W. E. 1990, MNRAS 245, 665

\end{thebibliography}
\end{document}